\documentclass[prl, twocolumn,preprintnumbers,amsmath,amssymb,showpacs,aps,prl]{revtex4}
\usepackage{mathrsfs}

\usepackage{graphicx}
\usepackage{dcolumn}
\usepackage{amsmath}
\usepackage{epsfig}
\usepackage{subfigure}
\usepackage{color}
\usepackage{epstopdf}

\begin{document}

\date{\today}

\title{Entanglement's Benefit Survives an Entanglement-Breaking Channel}%
\author{Zheshen Zhang}%
\email{zszhang@mit.edu}%
\author{Maria Tengner}
\author{Tian Zhong}
\author{Franco N. C. Wong}
\author{Jeffrey H. Shapiro}

\affiliation{Research Laboratory of Electronics, Massachusetts Institute of Technology,
77 Massachusetts Avenue, Cambridge, Massachusetts 02139, USA}%

\begin{abstract} 
Entanglement is essential to many quantum information applications, but it is easily destroyed by quantum decoherence arising from interaction with the environment.  We report the first experimental demonstration of an entanglement-based protocol that is resilient to loss and noise which destroy entanglement.  Specifically, despite channel noise 8.3 dB beyond the threshold for entanglement breaking, eavesdropping-immune communication is achieved between Alice and Bob when an entangled source is used, but no such immunity is obtainable when their source is classical. The results prove that entanglement can be utilized beneficially in lossy and noisy situations, i.e., in practical scenarios.
\end{abstract}

\maketitle

Entanglement is essential to many quantum information applications \cite{bennett93,bouwmeester97,ma12,kimble98,bennett92,mattle96,barreiro08,shor97,vandersypen01,politi09,ekert91}, but it is easily destroyed.  Quantum illumination (QI) \cite{lloyd08,tan08,guha09,shapiro09} is a radically different entanglement-based paradigm for bosonic channels: it thrives on entanglement-breaking loss and noise. For a given transmitter power, an initially entangled state's nonclassical correlation produces a classical state at the output of an entanglement-breaking channel whose correlation can greatly exceed what any classical input of the same power can yield through that channel.  This suggests that bosonic entanglement can be utilized advantageously in practical situations where it does not survive.  

First proposed to increase the signal-to-noise ratio (SNR) for detecting a weakly-reflecting target in the presence of strong background noise \cite{lloyd08,tan08,guha09}, quantum illumination was later shown, theoretically, to enable high data-rate classical communication that is immune to passive eavesdropping \cite{shapiro09}.  In the latter application, Alice and Bob use an entangled-state input for their data transfer. Eve, however, has no access to Alice's retained portion of the entangled state, so her eavesdropping performance is that of a classical-state input. The resulting disparity between Alice and Eve's performance---in bit-error rate (BER) and information received per transmitted bit---guarantees Alice and Bob's communication security. In this Letter we report the first experimental demonstration of QI's passive-eavesdropping immunity.   Aside from its relevance to secure communication, our experiment represents the first time that bosonic entanglement has yielded a strong performance benefit over an entanglement-breaking channel.  Thus it implies that the use of entanglement should \em not\/\rm\ be dismissed for environments in which it will be destroyed.  Moreover, unlike the recent experiment \cite{lopaeva13} reporting the target-detection advantage of photon-pair correlations, our eavesdropping-immune QI protocol \em requires\/\rm\ an initial state that is entangled.  Also, our communication protocol uses only one pulse to decode a bit, whereas target detection in \cite{lopaeva13} depends on the accumulation of enough data to accurately estimate a covariance.  

\begin{figure*}[hbt]
\includegraphics[width=5.5in]{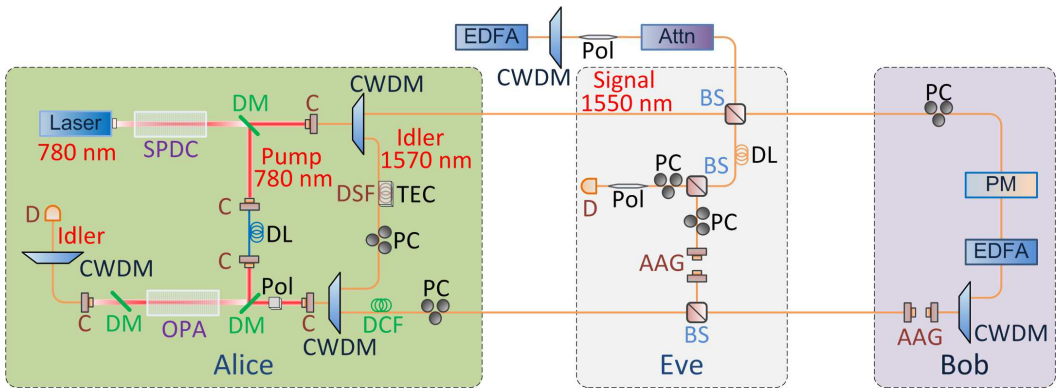}
\caption{\label{FigExp} (color online). Experiment setup. SPDC: spontaneous parametric downconverter; DM: dichroic mirror; C: collimator; CWDM: coarse  wavelength-division multiplexer; BS: beam splitter; Attn: attenuator; EDFA: erbium-doped fiber amplifier; DL: delay line; PC: polarization controller; PM: phase modulator; AAG:  adjustable air gap; Pol: polarizer; DCF: dispersion-compensating fiber; DSF: dispersion-shifted fiber; TEC: thermoelectric cooler; OPA: optical parametric amplifier; D: detector.} 
\end{figure*}

Our QI communication experiment is shown schematically  in Fig.~\ref{FigExp}. Alice prepares maximally-entangled signal and idler beams using a spontaneous parametric downconverter (SPDC), sending the signal to Bob and retaining the idler. Bob encodes his message bits at 500\,kbit/s by applying $0$ (message bit = 0) or $\pi$\,rad (message bit = 1) phase shifts on the signal he receives from Alice. Bob then intentionally breaks the signal-idler entanglement by passing his modulated signal light through an erbium-doped fiber amplifier (EDFA), whose amplified spontaneous emission (ASE) noise masks his bit stream from Eve.   Eve must rely on the joint classical-state light she has tapped from the Alice-to-Bob and Bob-to-Alice channels, while Alice combines her noisy returned light with her retained idler in a joint measurement to decode Bob's bit stream.  Consequently, QI makes Alice's cross-correlation signature between her retained and returned light beams far stronger than Eve's corresponding signature for her two tapped beams, even though Alice and Eve's receivers only have classical states at their disposal. 

In Fig.~\ref{FigExp}, Alice's SPDC uses a 20-mm type-0 phase-matched MgO-doped periodically-poled lithium niobate (MgO:PPLN) crystal that is continuous-wave (cw) pumped at 780\,nm, producing signal and idler outputs at 1550\,nm and 1570\,nm. A coarse wavelength division multiplexer (CWDM) separates the signal and idler and bandlimits them to 16\,nm ($W \approx 2\,$THz).  The $T = 2$\,$\mu$s bit duration at 500\,kbit/s then contains $M= TW \approx 4\times 10^6$ signal-idler mode pairs per information bit. At $\sim$135\,mW pump power, the SPDC generates a source brightness of $N_S = 0.001$ signal (and idler) photons per mode on average. Having $N_S \ll 1$  and $MN_S \gg 1$ are essential for QI.  The former ensures that Alice gets a much stronger correlation signature than Eve does, and the latter guarantees that Alice receives sufficient photons/bit to achieve a low error probability.  We typically operate at pump levels yielding $MN_S$ values of hundreds to thousands. 

Alice retains the idler in a spool of dispersion-shifted fiber, whose propagation delay matches the Alice-to-Bob-to-Alice delay seen by the signal beam. She sends her signal beam to Bob through a single-mode fiber (SMF) into which Eve has placed a 50-50 beam splitter.  Bob applies binary phase-shift keying (BPSK) modulation to the signal light he has received using a phase modulator driven by a pseudorandom bit sequence from a bit-error rate tester (BERT).  The modulated light is fed to an EDFA set to a measured gain $G_B \approx 1.34 \times 10^4$ whose ASE noise has per-mode average photon number $N_B \approx 1.46 \times 10^4$. A CWDM filter is used to bandlimit the ASE to the 16\,nm occupied by the signal and to attenuate the ASE within the idler spectral band by $\sim$30\,dB.  Complete suppression of the ASE noise outside of the signal band is achieved with a second CWDM in Alice's receiver (and with additional attenuation in Eve's receiver).  

Our QI protocol is intrinsically interferometric, so Bob uses a free-space delay line with $\sim$80\% efficiency to fine tune the timing between the signal and the idler paths. Dispersion in the SMFs connecting Alice and Bob broadens the SPDC's $\sim$0.22\,ps biphoton wave-function to $\sim$27\,ps. Thus Alice injects the light returned from Bob into $\sim$10\,m of dispersion-compensating fiber before combining it with her retained idler through a CWDM. The signal path sustains a measured channel loss of $\sim$16.4\,dB that includes SMF coupling loss, fiber-optic component insertion loss, and Eve's 50\% (10\%) tap placed before (after) Bob's apparatus. (Eve's 50\% tap of the Alice-to-Bob channel minimizes her BER when her receiver is ASE limited.  However, her removing more than 10\% of light from the Bob-to-Alice channel does not improve her BER, because her receiver is ASE limited even with the 10\% tap.)  Alice's idler suffers $\sim$4.1\,dB channel loss from SMF coupling and component insertion loss.

Alice decodes Bob's message bits by applying the returned and retained light to the signal and idler ports, respectively, of a low-gain optical parametric amplifier (OPA), and then doing direct detection on the OPA's idler-port output followed by matched filtering of the output current and threshold-decision logic. In the experiment, the returned and retained light are free-space coupled with the cw pump beam through a dichroic mirror to an OPA based on a 20-mm MgO:PPLN crystal.  The OPA converts the cross correlation between the phase-modulated signal and the retained idler into amplitude modulation in the output idler that can be sensed with direct detection. After the OPA, a dichroic mirror is used to remove the pump and the OPA's signal and idler outputs are coupled into an SMF and separated by a CWDM filter.  The separated idler is coupled into free space and detected by an avalanche photodiode (APD) setup that is 45\% efficient, when coupling and CWDM loss are combined with detector quantum efficiency. 

The APD's output current passes through a low-noise current amplifier, whose output is sent to an active high-pass filter, to reject dc, followed by a low-pass filter.   The sampled output from the second filter is supplied to a field programmable gate array (FPGA) that yields two outputs.  The FPGA program to produce the first output approximates the matched filter for a single bit, and it is subsequently dc shifted and amplified to transistor-transistor logic levels for BER measurements.  The second output provides a feedback signal to a lock-in amplifier that is part of a servo-control system (SCS) which stabilizes the relative phases between the OPA pump, Alice's retained idler, and the modulated light she receives from Bob.  The SCS also includes a slow thermal-control loop for Alice's fiber spool.  Typical incident power at the APD is approximately 10\,nW.  It is dominated by the signal-band ASE noise converted to the idler band by the OPA.  The OPA gain $G_A$ is kept low, $G_A-1\ll 1$, to prevent the ASE noise from overwhelming the amplitude modulation in the OPA's idler output. We implement Eve to demonstrate Alice's entangled-input QI performance advantage over what Eve achieves with her classical-state input. Eve decodes Bob's message by combining the light she has tapped from Alice and Bob's transmissions on an asymmetric beam-splitter, and then doing direct detection followed by matched filtering of the output current and threshold-decision logic.  

Both Alice and Eve's receivers are sub-optimal, in comparison with minimum error-probability quantum measurements, but both represent their best receivers for which explicit realizations are known. With these receivers Alice and Eve's BERs are given by (see Supplemental Material for details)
\begin{eqnarray}
\label{eqBERA_exp}
\text{BER}_A &=& Q\left(\frac{\sqrt{M}\zeta_A \sqrt{N_S(N_S+1)}}{\sigma_{A_+}^\text{tot}+\sigma_{A_-}^\text{tot}}\right)\\ 
\text{BER}_E &=& Q\left(\frac{\sqrt{M}\zeta_E N_S}{\sigma_{E_+}^\text{tot}+\sigma_{E_-}^\text{tot}}\right).
\end{eqnarray}
Here: $Q$ is the tail integral of the standard Gaussian probability density; $\zeta_A \sqrt{N_S(N_S+1)}$ ($\zeta_EN_S$) is the modulation-depth signature of Bob's message bit seen by Alice (Eve), where $\zeta_A$ ($\zeta_E$) is a transmission efficiency; and $ \sigma_{A_{\pm}}^{\text{tot}} $ ($ \sigma_{E_{\pm}}^{\text{tot}}  $) are Alice's (Eve's) per-mode noise standard deviations for bit values 0 and 1.  The transmission efficiencies include (see  Supplemental Material) the EDFA gain, channel loss, and an effective modulation-depth factor due to residual dispersion and less than optimal mode-pair  coupling into an SMF.  

\begin{figure}[hbt]
\includegraphics[scale = 0.4]{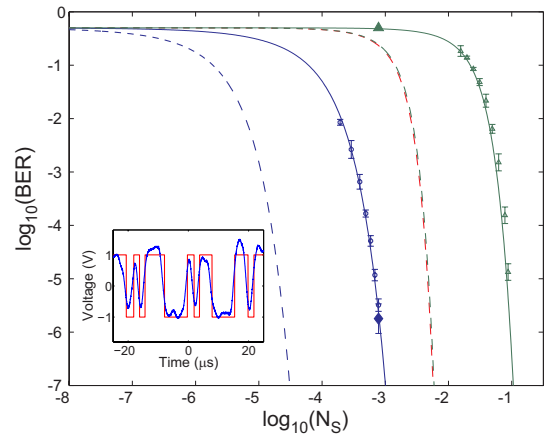}
\caption{\label{FigBER_vs_pumppower} BER$_A$ and BER$_E$ versus source brightness $N_S$ for 500\,kbit/s communication.  Inset: 25\,bits of OPA-receiver detector output (blue) and Bob's corresponding modulation waveform (red). See text for more information.} 
\end{figure}

The points in Fig.~\ref{FigBER_vs_pumppower} are Alice and Eve's measured BERs.  Each point is the average of 10 measurements---1\,Msample per measurement, which typically takes only a few seconds---with error bars indicating $\pm$1 standard deviation. At Alice's maximum SPDC output, Eve's measured modulation depth is insufficient to permit BER$_E$ measurements  that would reveal their $N_S$ dependence. Thus to demonstrate that scaling, we replace Alice's SPDC source with the attenuated, CWDM-filtered ASE from an EDFA, as shown in Fig.~\ref{FigExp}, whose flat-top spectrum thermal state mimics SPDC signal light, but easily produces tens of nW of power for Eve.  Eve stores the light split from the Alice-to-Bob channel using a fiber spool, which matches the total fiber length inside Bob's setup, and employs a free-space delay line  to fine tune the timing between the two paths. The free-space delay line's coupling is adjusted to optimize Eve's receiver and further suppress the out-of-band ASE noise.  Light she has split from the Alice-to-Bob and Bob-to-Alice channels are interfered on a 99-1 beam splitter, with 99\% of the power coming from the light taken from the Alice-to-Bob channel, and 1\% of the power coming from the light taken from the Bob-to-Alice channel. Eve uses a feedback arrangement similar to Alice's to stabilize her interferometer.  Eve obtains her BER values by measuring the combined light using the same APD detection setup employed by Alice. 

The dashed and solid blue curves in Fig.~\ref{FigBER_vs_pumppower} are theory for BER$_A$ when Alice uses a maximally-entangled SPDC  and an OPA receiver with gain $G_A - 1 = 1.86 \times 10^{-5}$.  The dashed blue curve shows Alice's performance when she has an ideal OPA receiver, viz., no loss of modulation depth due to residual dispersion or sub-optimal mode-pair coupling, unity detection efficiency, unity APD noise figure, no OPA pump-power fluctuations, and no electronics noise; the solid blue curve employs the experimentally-determined values for these receiver nonidealities.  The dashed red curve---virtually identical to the dashed green curve---assumes that Alice uses a classical-state source with maximally-correlated signal and idler and an ideal OPA receiver.  The gap between the dashed red and solid blue curves shows that Alice's performance using an SPDC source and imperfect OPA reception exceeds what can be achieved with that classical-state source and ideal OPA reception.  

The dashed and solid green curves in Fig.~\ref{FigBER_vs_pumppower} are theory for BER$_E$ when Alice uses a maximally-entangled SPDC source \em or\/\rm\ a maximally-correlated classical source and Eve employs an interference receiver.  The dashed curve assumes Eve's receiver is ideal; the solid green curve employs the experimentally-determined values for her receiver's nonidealities.  The near-identical nature of the dashed red and dashed green curves is a coincidental consequence of our experiment's parameter values.   

The blue circles in Fig.~\ref{FigBER_vs_pumppower} are measured BER$_A$ values under the operating conditions used to obtain the solid blue curve; they show our experimental results to be in excellent agreement with theory with no free parameters being adjusted.  The filled blue diamond in Fig.~\ref{FigBER_vs_pumppower}  is Alice's measured BER  at $N_S = 7.81 \times 10^{-4}$ when her OPA gain was increased to $G_A-1 = 2.48 \times 10^{-5}$, and the filled green triangle above it is the measured BER$_E$ (Alice's SPDC is used for this measurement).  These two points represent our secure-communication operating point at which BER$_A$ = 1.78 $\times 10^{-6}$ and BER$_E \approx 0.5$.  The inset provides a visual indication of Alice's low-BER performance by overlaying 25 bits of her  receiver output (blue), with the dc level removed, on Bob's corresponding modulation waveform (red), which is scaled to match the data's peak-to-peak range.  The joint state of Alice's returned and retained beams, conditioned on Bob's BPSK value, is zero-mean and Gaussian.  Hence it becomes classical  (see Supplemental Material) when $N_B \ge N_B^{\rm thresh} = 2.14 \times 10^3$, so our measured $N_B = 1.46\times 10^4$ is 8.3\,dB above the threshold for classicality.   

The disparity between Alice and Eve's BERs at the secure-communication operating point, the $N_S$ gap between the dashed red and solid blue theory curves at the same BER values, and $N_B$'s exceeding $N_B^{\rm thresh}$ by 8.3\,dB confirm the essential feature of quantum illumination that is exploited here in a communication setting: a large performance gap between an entangled-state input and a classical-state input in a lossy and noisy channel.  

The open green triangles in Fig.~\ref{FigBER_vs_pumppower}  were obtained using attenuated ASE from an EDFA source in lieu of light from a downconverter.  They show our measurements to be in excellent agreement with theory with no free parameters being adjusted.  The $N_S$ gap between the blue circles and the green triangles in Fig.~\ref{FigBER_vs_pumppower} at the same BER values quantifies Alice and Bob's entanglement-derived communication advantage when Alice and Eve both use realistic receivers. 

At this point it is instructive to consider what would happen had Alice used a cw laser, instead of an SPDC source, and did homodyne detection on the returned light from Bob.  That receiver is ASE limited---and essentially quantum optimal---with an error-exponent that is 3\,dB inferior to that of an ideal OPA receiver for the same transmitted power, so it might seem to provide a much easier route to passive eavesdropping immunity.  Such, however, is not the case:  when Eve taps the Alice-to-Bob and Bob-to-Alice links and uses her own homodyne receiver, it too is ASE limited and its BER is the same as Alice's.   So Alice's initial broadband signal-idler entanglement is absolutely essential to her obtaining immunity to Eve's passive eavesdropping, because Eve can then only employ the broadband thermal light she taps from Alice and its modulated version that she captures from Bob.

Our measurements show Alice enjoying five orders-of-magnitude error-probability advantage over Eve---when both use realistic receivers---at the secure-communication operating point.  Moreover, if Bob's bit sequence is equally-likely and statistically independent then:  (1) Alice receives very close to one bit of information for each bit that Bob has transmitted; and (2) Eve receives nearly zero information about each of Bob's bits at this operating point.  However, both Alice and Eve's receivers are limited more by their technical noise than the ASE-generated noise that sets their ultimate sensitivity:  at the secure-communication operating point, Alice's ASE-generated variance is only 26.7\% (25.1\%) for $0$\,rad modulation ($\pi$\,rad modulation) of her total noise variance.  


It is useful to evaluate the QI performance gap in a different way: Alice's information advantage over Eve. The two panels in Fig.~\ref{InfoAdvantage} display Alice's Shannon information $I_{AB}$, an upper bound on Eve's Holevo information $\chi_{EB}^{\rm UB}$, and a lower bound on Alice's information advantage $\Delta I_{AB}^{\rm LB} \equiv I_{AB} - \chi^{\rm UB}_{EB}$.  (See Supplemental Material for details.)   In the top panel, Alice's Shannon information is computed using the error probabilities from the solid blue curve in Fig.~\ref{FigBER_vs_pumppower}, while the upper bound on Eve's Holevo information is the most she could learn about an infinitely long sequence of Bob's bits from an optimum collective quantum measurement on the light she extracts from the Alice-to-Bob and Bob-to-Alice channels.  Here we see that with Alice using her imperfect OPA receiver she can get  up to 0.8\,bits of information advantage, per Bob's transmitted bit, over Eve's optimum collective quantum measurement.  
\begin{figure}[htb]
\label{InfoAdvantage}
\includegraphics[scale = 0.45]{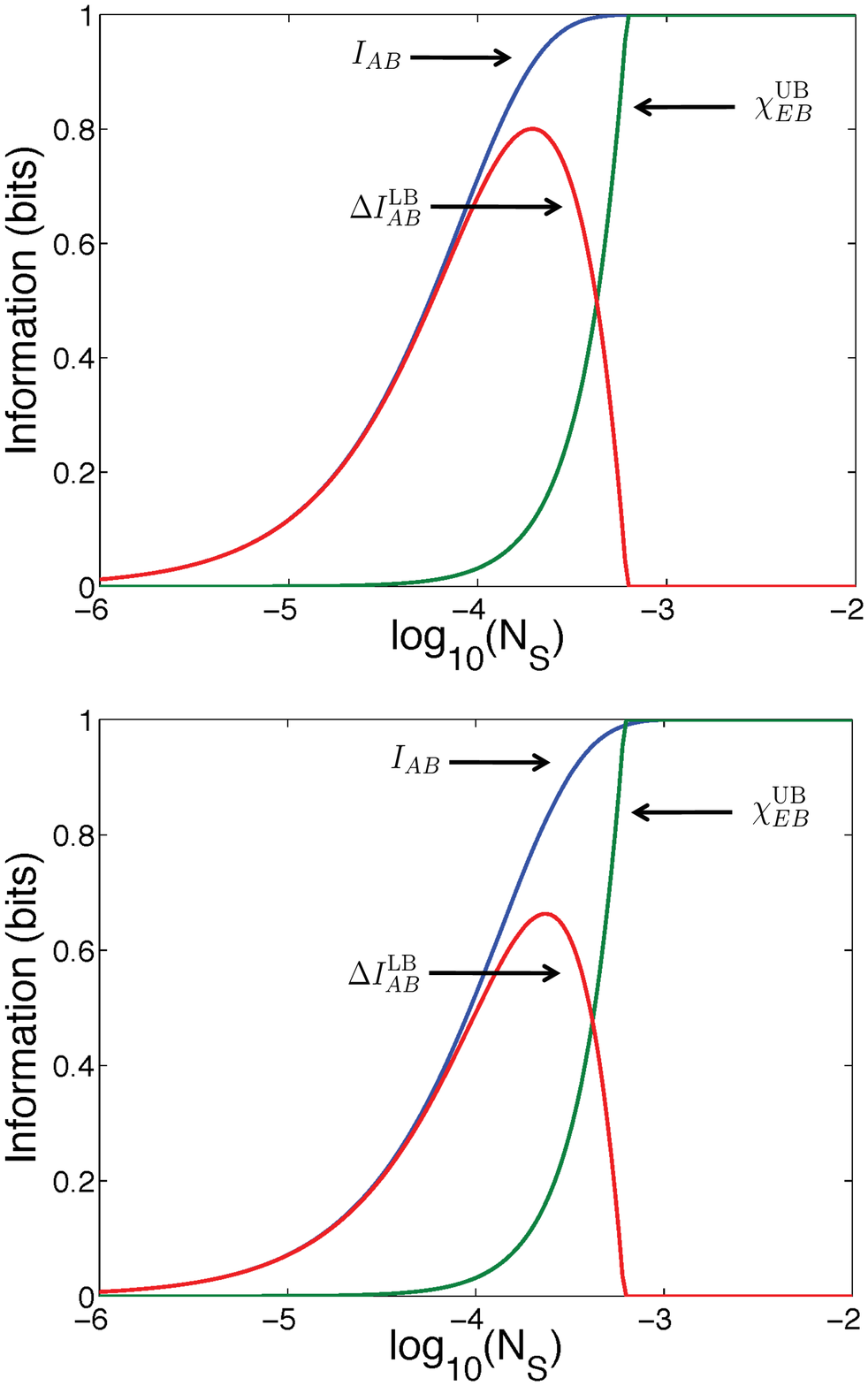}
\caption{\label{InfoAdvantage} (color online).  Alice's Shannon information, an upper bound on Eve's Holevo information, and a lower bound on Alice's information advantage, versus Alice's source brightness $N_S$.  The top panel computes Alice's Shannon information using the error probabilities from the solid blue curve in Fig.~\ref{FigBER_vs_pumppower}.  The bottom panel computes Alice's Shannon information when Eve takes 90\% of the Bob-to-Alice light instead of the 10\% she receives in the top panel.} 
\end{figure}

A more convincing demonstration that quantum illumination offers Alice a significant information advantage is shown in the bottom panel of Fig.~\ref{InfoAdvantage}.  Here we have changed the beam splitter that Eve uses to tap the Bob-to-Alice transmission to have 10\% transmissivity instead of the 90\% value used in our experiment.  Under these conditions, Eve gets the same amount of Alice's light that Bob does and nine times the amount of Bob's light that Alice does.  Nevertheless, the upper bound on her Holevo information is unchanged from what is seen in the top panel.  This is because her received BPSK modulation depth and the standard deviation of the ASE-generated noise that dominates the fluctuations in the jointly-Gaussian state she receives both increase in proportion to the fraction of the light she taps.  Alice's BER, however, is somewhat degraded by reducing the amount of light she gets from Bob, because of her receiver's technical noise.  Consequently, her information advantage over Eve's optimum collective quantum measurement now peaks at 0.66\,bits/bit.  

Our experiment demonstrates the QI-communication protocol's immunity to \em passive\/\rm\ eavesdropping that was predicted, theoretically, in \cite{shapiro09}.   That reference already noted that this protocol is vulnerable to \em active\/\rm\ attacks, in which Eve injects her own light into Bob's terminal.  Indeed, Eve could inject her own SPDC signal light into Bob, retaining her source's idler light for use with the light she taps from the Bob-to-Alice channel.   Steps that could be taken to ward off active attacks were suggested in \cite{shapiro09}, and initial analyses of many of these approaches have been performed \cite{xu11}, but more work is needed on defeating active attacks.  

In conclusion, we have demonstrated that the benefits of bosonic entanglement  can be reaped over an entanglement-breaking channel. Our QI protocol experimentally achieved more than five orders-of-magnitude BER advantage for Alice over a passive eavesdropping Eve when both use realistic receivers.  Furthermore, based on the excellent fit between our data and theory, we claim that Alice enjoys an information advantage that can exceed 0.6\,bits per Bob's transmitted bit over the optimum collective quantum Eve when Eve taps 50\% of the Alice's transmission and 90\% of Bob's transmission. These advantages are consequences of Alice's initial broadband signal-idler entanglement, because they disappear when Alice's signal-idler phase-sensitive cross correlation is classical, or when Alice uses a laser source.  The meta-lesson of our experiment is as follows:  the performance disparity between a QI entangled-state input and a thermal-state input in communication over a lossy, noisy bosonic channel suggests that entanglement may be utilized to good effect in other entanglement-breaking environments.

The authors thank Eric Dauler for the loan of the bit-error rate tester.  This research was supported by an ONR Basic Research Challenge grant.

\newpage

\widetext
\begin{center}{\Large\bf Supplemental Material}
\end{center}%

\section{I.\,\,  The communication model}
\begin{figure*}[hbt]
\includegraphics[width=4.5in]{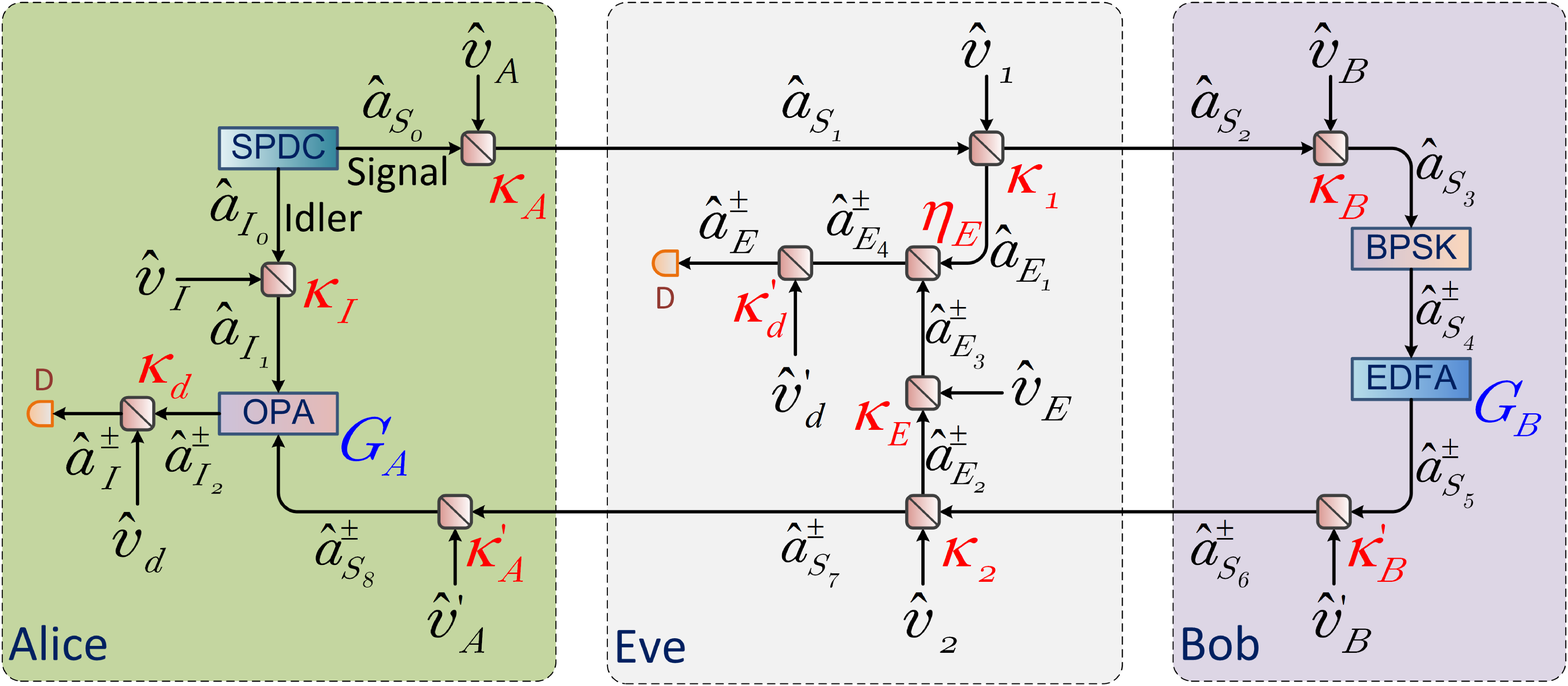}
\caption{\label{FigModel} Conceptual model for secure communication based on quantum illumination.  SPDC:  spontaneous parametric downconverter.  BPSK:  binary phase-shift keying.  EDFA:  erbium-doped fiber amplifier.  D:  detector.  OPA:  optical parametric amplifier.} 
\end{figure*}

Figure~\ref{FigModel} shows a conceptual schematic for our experiment.  The $\{\hat{a}_{S_i}\}$ and $\{\hat{a}_{I_i}\}$ are signal and idler annihilation operators for one of the $M$ independent, identically-distributed, signal-idler mode pairs produced by Alice's spontaneous parametric downconverter (SPDC). The annihilation operators $\hat{a}^{\pm}_I$ and $\hat{a}^\pm_E$ are associated with the modes Alice and Eve measure, respectively, from a single signal-idler mode pair, where the $\pm$ superscripts here and elsewhere in the figure represent Bob's $\pm 1$ (0 or $\pi$\,rad) binary phase-shift keying (BPSK) modulation.  The transmissivities $\kappa_I$, $\kappa_A$, $\kappa_1$, $\kappa_B$, $\kappa'_B$, $\kappa_2$, $\kappa'_A$, $\kappa_d$, $\kappa_E$, and $\kappa_d'$ model losses---including detector quantum efficiencies---in Alice, Bob, and Eve's terminals and the propagation channels that link them, with the annihilation operators $\{\hat{v}_I,\hat{v}_A, \hat{v}_1, \hat{v}_B, \hat{v}'_B,  \hat{v}_2,\hat{v}'_A, \hat{v}_d,\hat{v}_E,\hat{v}'_d\}$ representing vacuum-state auxiliary modes. Bob's erbium-doped fiber amplifier (EDFA) has gain $G_B$ and Alice's optical parametric amplifier (OPA) has gain $G_A$.  Eve chooses the transmissivities $\kappa_E$ and $\eta_E $ to minimize her bit error rate (BER).  Additional details are as follows.

\noindent{\bf Propagation and detection loss.}  Propagation and detection loss are accounted for, in Fig.~\ref{FigModel}, by beam splitters.  Each of these is modeled by the annihilation-operator relation
\begin{equation}
\hat{a}_{\rm out} = \sqrt{\kappa}\,\hat{a}_{\rm in} + \sqrt{1-\kappa}\,\hat{v},
\end{equation}
in terms of the input $\hat{a}_{\rm in}$, the output $\hat{a}_{\rm out}$, and a vacuum-state auxiliary mode $\hat{v}$, where $0 < \kappa <1$ is the beam-splitter's transmissivity.\\

\noindent{\bf BPSK model.}  Bob's BPSK modulator is modeled, for each signal mode, by the annihilation-operator 
relation
\begin{equation}
\hat{a}_{S_4}^\pm = \pm\hat{a}_{S_3},
\end{equation}
where $\pm$ denotes the common bit value applied to \em all\/\rm\ $M$ signal modes.\\ 

\noindent{\bf EDFA theory.}  Bob's EDFA is modeled, for each signal mode, by the annihilation-operator relation
\begin{equation}
\hat{a}_{S_5}^{\pm} = \sqrt{G_B}\,\hat{a}_{S_4}^\pm + \sqrt{G_B -1}\,\hat{a}_{\rm SE}^\dagger,
\end{equation}
in terms of the amplifier's gain $G_B$ and a thermal-state spontaneous-emission mode $\hat{a}_{\rm SE}$; the average photon-number per mode of amplified spontaneous emission (ASE) at the amplifier's output is $N_B = (G_B-1)(\langle \hat{a}_{\rm SE}^\dagger\hat{a}_{\rm SE}\rangle+1)$.\\

\noindent{\bf OPA theory.}  Alice's OPA is modeled, for each returned-retained mode pair, by the annihilation-operator relation
\begin{equation}
\hat{a}_{I_2}^{\pm} = \sqrt{G_A}\,\hat{a}_{I_1} + \sqrt{G_A-1}\,\hat{a}_{S_8}^{\pm\dagger}.
\end{equation}

\section{II.\,\, Mode-pair photon statistics}
A single signal-idler mode pair $\{\hat{a}_{S_0},\hat{a}_{I_0}\}$ from Alice's continuous-wave spontaneous parametric downconverter is in the maximally-entangled, zero-mean, jointly-Gaussian state $\hat{\rho}_{SI}$ whose density operator is completely characterized by the covariance matrix
$\Gamma_{SI} \equiv \langle[\begin{array}{cccc}\hat{a}_{S_0}^\dagger & \hat{a}_{I_0}^\dagger & \hat{a}_{S_0} & \hat{a}_{I_0}\end{array}]^T[\begin{array}{cccc}\hat{a}_{S_0} & \hat{a}_{I_0} & \hat{a}_{S_0}^\dagger & \hat{a}_{I_0}^\dagger\end{array}]\rangle$ given by
\begin{equation}
\label{eqcovariance}
\Gamma_{SI} = \left[\begin{array}{cccc} N_S & 0 & 0 & \sqrt{N_S(N_S+1)}\\[.05in] 
0 & N_S & \sqrt{N_S(N_S+1)} & 0 \\[.05in]
0 & \sqrt{N_S(N_S+1)} & N_S + 1 & 0 \\[.05in]
\sqrt{N_S(N_S+1)} & 0 & 0 & N_S + 1\end{array}\right],
\end{equation}
where $N_S$ is the source brightness (average number of signal-idler photon pairs per mode).  
This density operator plus the input-output operator evolutions associated with the beam splitters, phase modulator, and optical amplifiers shown in the quantum illumination conceptual model from Fig.~\ref{FigModel} determine Alice and Eve's single-mode photon-counting statistics, namely those of the $\hat{a}^{\pm\dagger}_I\hat{a}^\pm_I$ and $\hat{a}^{\pm\dagger}_E\hat{a}^\pm_E$ operators, respectively.   
Because all of the preceding evolutions are linear, with the beam splitter's free inputs being in their vacuum states and the EDFA's free input being in a thermal state, it follows that the $\hat{a}^\pm_I$ and $\hat{a}^\pm_E$ modes are in zero-mean Gaussian states.  Hence their photon-counting statistics are Bose-Einstein, with variances that satisfy
\begin{equation}
\sigma^2_{A_\pm} = N^\pm_A(N^\pm_A + 1),
\end{equation}
and
\begin{equation}
\sigma^2_{E_\pm} = N^\pm_E(N^\pm_E + 1),
\end{equation}
in terms of their average photon numbers $N^\pm_A \equiv \langle \hat{a}^{\pm\dagger}_I\hat{a}^\pm_I\rangle$ and $N^\pm_E \equiv \langle \hat{a}^{\pm\dagger}_E\hat{a}^\pm_E\rangle$.  It is now simple to back-propagate along the chain of linear evolutions to obtain explicit results for these average photon numbers.  

For Alice, including the effect of imperfect dispersion compensation and allowing for sub-optimal mode-pair coupling, we have that
\begin{eqnarray}
N_A^\pm &=& \kappa_d\langle \hat{a}^{\pm\dagger}_{I_2}\hat{a}^\pm_{I_2}\rangle\\
&=& \kappa_d\langle(\sqrt{G_A}\,\hat{a}_{I_1} + \sqrt{G_A-1}\,\hat{a}^{\pm\dagger}_{S_8})^\dagger(\sqrt{G_A}\,\hat{a}_{I_1} + \sqrt{G_A-1}\,\hat{a}^{\pm\dagger}_{S_8})\rangle\\ 
&=& \kappa_dG_A\langle\hat{a}^\dagger_{I_1}\hat{a}_{I_1}\rangle + \kappa_d(G_A-1) + \kappa_d(G_A-1)\langle\hat{a}^{\pm\dagger}_{S_8}\hat{a}^\pm_{S_8}\rangle  \nonumber\\
&& \pm\,\,2\eta^A_d \kappa_d\sqrt{G_A(G_A-1)}\,{\rm Re}(\langle \hat{a}_{I_1}\hat{a}^\pm_{S_8}\rangle),
\end{eqnarray}
where
\begin{equation}
\langle \hat{a}_{I_1}^\dagger\hat{a}_{I_1}\rangle = \kappa_I\langle \hat{a}_{I_0}^\dagger\hat{a}_{I_0}\rangle = \kappa_IN_S,
\end{equation}
\begin{eqnarray}
\langle\hat{a}^{\pm\dagger}_{S_8}\hat{a}^\pm_{S_8}\rangle &=& 
\kappa_A'\kappa_2\kappa_B'\langle\hat{a}^{\pm\dagger}_{S_5}\hat{a}^\pm_{S_5}\rangle\\ 
&=& \kappa_A'\kappa_2\kappa_B'G_B\langle\hat{a}^{\pm\dagger}_{S_4}\hat{a}^\pm_{S_4}\rangle + \kappa_A'\kappa_2\kappa_B'N_B\\ 
&=& \kappa_A'\kappa_2\kappa_B'G_B\kappa_B\kappa_1\kappa_A\langle\hat{a}^\dagger_{S_0}\hat{a}_{S_0}\rangle + \kappa_A'\kappa_2\kappa_B'N_B\\ 
&=& \kappa_A'\kappa_2\kappa_B'G_B\kappa_B\kappa_1\kappa_A N_S + \kappa_A'\kappa_2\kappa_B'N_B,
\end{eqnarray}
\begin{eqnarray}
\langle \hat{a}_{I_1}\hat{a}^\pm_{S_8}\rangle &=& \sqrt{\kappa_I\kappa_A'\kappa_2\kappa_B'}\,\langle \hat{a}_{I_0}\hat{a}^\pm_{S_5}\rangle \\ 
&=& \sqrt{\kappa_I\kappa_A'\kappa_2\kappa_B'G_B}\,\langle \hat{a}_{I_0}\hat{a}^\pm_{S_4}\rangle\\ 
&=& \pm\sqrt{\kappa_I\kappa_A'\kappa_2\kappa_B'G_B\kappa_B\kappa_1\kappa_A}\,\langle \hat{a}_{I_0}\hat{a}_{S_0}\rangle\\ 
&=& \pm\sqrt{\kappa_I\kappa_A'\kappa_2\kappa_B'G_B\kappa_B\kappa_1\kappa_A N_S(N_S+1)},
\end{eqnarray} 
$d_A$ is her ideal receiver's modulation depth, and $\eta^A_d$ is its modulation-depth efficiency factor.  
Combining the preceding results we obtain,
\begin{eqnarray}
N_A^\pm &=& \kappa_d G_A \kappa_I N_S + \kappa_d (G_A-1) + \kappa_d(G_A-1)\kappa_A' \kappa_2 \kappa_B' G_B \kappa _B \kappa_1 \kappa_A N_S + \kappa_d(G_A-1)\kappa_A' \kappa_2\kappa_B' N_B \nonumber \\
& & \pm\,\,2\eta^A_d \kappa_d\sqrt{G_A(G_A-1)\kappa_I\kappa_A'\kappa_2\kappa_B'G_B\kappa_B\kappa_1\kappa_AN_S(N_S+1)},
\end{eqnarray}
from which it follows that
\begin{equation}
\eta^A_d d_A \equiv N^+_A - N^-_A = 4\eta^A_d \kappa_d\sqrt{G_A(G_A-1)\kappa_I\kappa_A'\kappa_2\kappa_B'G_B\kappa_B\kappa_1\kappa_AN_S(N_S+1)}.
\end{equation}
In the Letter, we have used $\zeta_A\sqrt{N_S(N_S+1)} \equiv \eta^A_dd_A$.  

Similar to what we did in the previous paragraph, we have for Eve that
\begin{eqnarray}
N_E^\pm &=&\kappa'_d \langle \hat{a}^{\pm\dagger}_{E_4}\hat{a}^{\pm}_{E_4}\rangle \\ 
&=& \kappa'_d\langle (\sqrt{\eta_E}\,\hat{a}_{E_1} + \sqrt{1-\eta_E}\,\hat{a}_{E_3}^\pm)^\dagger
(\sqrt{\eta_E}\,\hat{a}_{E_1} + \sqrt{1-\eta_E}\,\hat{a}_{E_3}^\pm)\\
&=& \kappa'_d\eta_E\langle\hat{a}^\dagger_{E_1}\hat{a}_{E_1}\rangle + \kappa'_d (1-\eta_E)\langle \hat{a}^{\pm\dagger}_{E_3}\hat{a}^\pm_{E_3}\rangle + 2\eta^E_d \kappa'_d\sqrt{\eta_E(1-\eta_E)}\,{\rm Re}(\langle \hat{a}_{E_1}^\dagger\hat{a}^\pm_{E_3}\rangle)\\
&=& \kappa'_d \eta_E(1-\kappa_1)\kappa_A\langle \hat{a}_{S_0}^\dagger\hat{a}_{S_0}\rangle + \kappa'_d (1-\eta_E)\kappa_E(1-\kappa_2)\kappa_B'\langle\hat{a}^{\pm\dagger}_{S_5}\hat{a}^\pm_{S_5}\rangle \nonumber \\ && +\,\, 
2\eta^E_d \kappa'_d \sqrt{\eta_E(1-\eta_E)(1-\kappa_1)\kappa_A\kappa_E(1-\kappa_2)\kappa_B'}\,{\rm Re}(\langle \hat{a}^\dagger_{S_0}\hat{a}^\pm_{S_5}\rangle),
\end{eqnarray}
where
\begin{equation}
\langle\hat{a}^{\pm\dagger}_{S_5}\hat{a}^\pm_{S_5}\rangle = G_B\kappa_B\kappa_1\kappa_A N_S + N_B,
\end{equation}
\begin{equation}
\langle \hat{a}^\dagger_{S_0}\hat{a}^\pm_{S_5}\rangle = \pm\sqrt{G_B\kappa_B\kappa_1\kappa_AN_S^2},
\end{equation}
$d_E$ is her ideal receiver's modulation depth, and $\eta^E_d$ is its modulation-depth efficiency factor.  
Combining the preceding results we get
\begin{eqnarray}
N^\pm_E &=& \kappa'_d \eta_E(1-\kappa_1)\kappa_AN_S + \kappa'_d (1-\eta_E)\kappa_E(1-\kappa_2)\kappa_B' G_B\kappa_B\kappa_1\kappa_AN_S + \kappa'_d (1-\eta_E)\kappa_E(1-\kappa_2)\kappa_B'N_B \nonumber \\ 
&& \pm\,\,2\eta^E_d \kappa'_d\sqrt{\eta_E(1-\eta_E)(1-\kappa_1)\kappa_A^2\kappa_E(1-\kappa_2)\kappa_B'G_B\kappa_B\kappa_1N_S^2},
\end{eqnarray}
from which it follows that
\begin{equation}
\eta^E_d d_E \equiv N^+_E - N^-_E = 4\eta^E_d\kappa'_d \sqrt{\eta_E(1-\eta_E)(1-\kappa_1)\kappa_A^2\kappa_E(1-\kappa_2)\kappa_B'G_B\kappa_B\kappa_1N_S^2}.
\end{equation}
In the Letter, we have used $\zeta_EN_S \equiv \eta^E_dd_E$.  

In order to evaluate our bound, developed below, on Eve's Holevo information, we need expressions for $N_{E_1}\equiv \langle \hat{a}^\dagger_{E_1}\hat{a}_{E_1}\rangle$, $N_{E_3}\equiv \langle \hat{a}^{\pm\dagger}_{E_3}\hat{a}^\pm_{E_3}\rangle$, and $C^\pm_{E_1E_3} \equiv \langle \hat{a}_{E_1}^\dagger\hat{a}^\pm_{E_3}\rangle$.  These are obtained as follows.  For $N_{E_1}$ we have that
\begin{equation}
N_{E_1} = (1-\kappa_1)\kappa_A\langle \hat{a}^\dagger_{S_0}\hat{a}_{S_0}\rangle = (1-\kappa_1)\kappa_A N_S.
\end{equation}
For $N_{E_3}$ we have that
\begin{equation}
N _{E_3} = (1-\kappa_2)\kappa_B'\langle\hat{a}^{\pm\dagger}_{S_5}\hat{a}^\pm_{S_5}\rangle = 
(1-\kappa_2)\kappa_B' G_B\kappa_B\kappa_1\kappa_A N_S + (1-\kappa_2)\kappa_B'N_B.
\end{equation}
For $C^\pm_{E_1E_3}$ we have that
\begin{equation}
C^\pm_{E_1E_3} = \sqrt{\kappa_A(1-\kappa_1)(1-\kappa_2)\kappa_B'}\,\langle \hat{a}^\dagger_{S_0}\hat{a}^\pm_{S_5}\rangle =
\pm\sqrt{\kappa_A^2(1-\kappa_1)(1-\kappa_2)\kappa_B'G_B\kappa_B\kappa_1N_S^2}.
\end{equation}

\section{III.\,\, Alice and Eve's per-mode noise variances}
The per-mode conditional variances for Alice and Eve's ideal receivers when Bob's BSPK modulation is 0\,rad ($+$) or $\pi$\,rad ($-$) are the Bose-Einstein variances $\sigma^2_{A_\pm}$ and $\sigma^2_{E_\pm}$, respectively.  However, Alice and Eve's avalanche photodiodes have noise figure $F_{\rm APD}$, their detection systems have electronics noise with per-mode variance $\sigma_D^2$, and Alice's OPA receiver has a per-mode conditional variance $\sigma^2_{P_\pm} = 0.2\sigma^2_{A_\pm}$ arising from pump-power fluctuations.  Thus their total per-mode conditional variances are
\begin{equation}
\sigma^{{\rm tot}2}_{A_\pm} = F_{\rm APD}(\sigma^2_{A_\pm} + \sigma^2_{P_\pm}) + \sigma^2_D
\quad\mbox{and}\quad  
\sigma^{{\rm tot}2}_{E_\pm} = F_{\rm APD}\sigma^2_{A_\pm} + \sigma^2_D.
\end{equation}

\section{IV.\,\, Parameter values}
Table~\ref{table} lists experimentally-determined parameter values for the experiment that yielded the data reported as blue circles and green triangles in the Letter's Fig.~2.
\begin{table}[hbt]
\begin{tabular}{|c|c||c|}\hline
Parameter & Symbol & Value\\ \hline
Alice's fluorescence bandwidth & $W$ & 2\,THz\\
Alice's signal transmissivity & $\kappa_A$ & 0.74 \\
Alice-to-Bob transmissivity & $\kappa_1$ & 0.50\\
Bob's pre-EDFA transmissivity& $\kappa_B$ & 0.43 \\ 
Bob's BPSK modulation rate & $R$ & 500\,kbit/s\\ 
Bob's BSPK bit duration & $T$ & 2\,$\mu$s\\
Bob's EDFA gain & $G_B$ & $1.34\times 10^4$\\ 
Bob's EDFA per-mode ASE & $N_B$ & $1.46\times 10^4$\\ 
Bob's post-EDFA transmissivity & $\kappa_B'$ & 0.39\\
Bob-to-Alice transmissivity & $\kappa_2$ & 0.90\\ 
Alice's return transmissivity & $\kappa_A'$ & 0.41\\
Alice's idler transmissivity & $\kappa_I$ & 0.39\\ 
Alice's OPA gain -1 & $G_A-1$ & $1.86\times 10^{-5}$\\ 
Alice's detection efficiency & $\kappa_d$ & 0.45\\
Alice's modulation-depth efficiency & $\eta^A_d$ & 0.52\\ 
Alice's number of mode pairs & $M$ & $4\times 10^6$\\ 
Eve's return transmissivity & $\kappa_E$ &  0.013 \\ 
Eve's mixing transmissivity & $\eta_E$ & 0.99 \\ 
Eve's detection efficiency & $\kappa_d'$ & 0.5\\
Eve's modulation-depth efficiency & $\eta^E_d$ & 0.17\\ 
APD noise figure & $F_{\rm APD}$ & 3.0\\
Electronics-noise variance & $\sigma_D^2$ & $6.0 \times 10^{-3}$\\ \hline
\end{tabular}
\caption{Experimentally-determined parameter values for the experiment that yielded the BER data reported in the Letter's Fig.~2.}
\label{table}
\end{table}

\section{V.\,\, Bit-error rate for Gaussian-distributed observations}
Suppose that we observe a continuous random variable $x$, whose statistics depend on the discrete random-variable message $y$ that is equally likely to be $+1$ or $-1$.  In particular, let $x$ be Gaussian distributed with mean value $m_+$ and variance $\sigma^2_+$ when $y = +1$, and let $x$ be Gaussian distributed with mean value $m_-$ and variance $\sigma^2_-$ when $y = -1$, where $m_+ > m_-$ and $\sigma^2_+ > \sigma^2_-$.   With the appropriate identifications for the conditional means and variances, this setup encompasses both Alice and Eve's receiver statistics.

We will assume that the receiver employs a threshold-test decision rule,
\begin{equation}
x \begin{array}{c} \mbox{\scriptsize say $y=+1$ sent} \\ \ge \\ < \\ \mbox{\scriptsize say $y=-1$ sent} \end{array}\gamma,
\end{equation}
where the threshold, $\gamma$, is chosen to equalize the false-alarm and miss probabilities,
i.e.,
\begin{equation}
P_F \equiv \Pr(\,\mbox{say $y=+1$ sent}\mid \mbox{$y=-1$ sent}\,),
\end{equation}
and
\begin{equation}
P_M \equiv \Pr(\,\mbox{say $y=-1$ sent}\mid \mbox{$y=+1$ sent}\,).
\end{equation}
The threshold that yields this equality is
\begin{equation}
\gamma = \frac{m_+\sigma_- + m_-\sigma_+}{\sigma_+ + \sigma_-},
\end{equation}
which leads to 
\begin{equation}
P_F = P_M = Q\!\left(\frac{m_+ - m_-}{\sigma_+ + \sigma_-}\right).
\end{equation}
For the bit-error rate we thus get
\begin{equation}
{\rm BER} = \Pr(y = +1)P_M + \Pr(y=-1)P_F = Q\!\left(\frac{m_+ - m_-}{\sigma_+ + \sigma_-}\right).
\end{equation}
Moreover, because $P_F = P_M$ the binary communication channel whose input is $y$ and whose output is the result of the preceding threshold test is a binary symmetric channel.  

\section{VI.\,\, Alice and Eve's bit error rates}
Applying the central limit theorem to the sum of $M\gg 1$ conditionally independent (given Bob's BPSK modulation), identically distributed mode-pair contributions that comprise Alice and Eve measurements leads to a model of the form considered in the preceding section with
\begin{equation}
m_\pm \equiv M\bar{m}_A \pm M\zeta_A\sqrt{N_S(N_S+1)} /2\quad\mbox{and}\quad
\sigma^2_\pm \equiv M \sigma^{2{\rm tot}}_{A_\pm} 
\end{equation}
for Alice when she uses her SPDC source, and
\begin{equation}
m_\pm \equiv M\bar{m}_E \pm M\zeta_EN_S/2 \quad\mbox{and}\quad
\sigma^2_\pm \equiv M\sigma^{2{\rm tot}}_{E_\pm} 
\end{equation}
for Eve, where the constants $\bar{m}_A$ and $\bar{m}_E$ do not affect the BERs.  The Letter's Eqs.~(1) and (2) then follow immediately from the preceding BER theory for Gaussian-distributed observations.  When Alice uses a source that produces $M$ statistically independent, identically distributed, zero-mean, jointly-Gaussian signal-idler mode pairs whose phase-sensitive cross correlation is at the classical limit for the $N_S$ source brightness, her BER follows from the Gaussian-observation theory but with
\begin{equation}
m_\pm \equiv M\bar{m}_A \pm M\zeta_AN_S/2 \quad\mbox{and}\quad
\sigma^2_\pm \equiv M \sigma^{2{\rm tot}}_{A_\pm}. 
\end{equation}

\section{VII.\,\, Alice's Shannon information and Eve's Holevo information}  
Alice's Shannon information (in bits) is given by
\begin{equation}
I_{AB} = \sum_{a,b}\Pr(a,b)\log_2\!\left(\frac{\Pr(b\mid a)}{\Pr(b)}\right),
\end{equation}
where $b = 0,1$ and $a = 0,1$ are the values of Bob's transmitted and Alice's decoded bits.
Taking Bob's bit to be equally likely 0 or 1, and using the fact that the Bob-to-Alice channel is binary symmetric---i.e., Alice's false-alarm and miss probabilities both equal the error probability from the Letter's Eq.~(1)---we get
\begin{eqnarray}
I_{AB} &=& 1 - {\rm BER}_A\log_2({\rm BER}_A)\\
&-& (1-{\rm BER}_A)\log_2(1-{\rm BER}_A).
\end{eqnarray}

Eve collects a set of $M$ conditionally independent (given Bob's modulation), identically-distributed mode pairs from the Alice-to-Bob and Bob-to-Alice channels.  Denoting the annihilation operators for a generic mode pair from this set by $\{\hat{a}_{E_1},\hat{a}^{\pm}_{E_2}\}$, we have that each mode pair, given Bob's modulation, is in a zero-mean, jointly-Gaussian state whose density operator $\hat{\rho}^\pm_{E_1E_2}$ is completely characterized by the covariance matrix $\Gamma^{\pm}_{E_1E_2} \equiv \langle[\begin{array}{cccc}\hat{a}_{E_1}^\dagger & \hat{a}_{E_2}^{\pm\dagger} & \hat{a}_{E_1} & \hat{a}_{E_2}^\pm\end{array}]^T[\begin{array}{cccc}\hat{a}_{E_1} & \hat{a}_{E_2}^\pm & \hat{a}_{E_1}^\dagger & \hat{a}_{E_2}^{\pm\dagger}\end{array}]\rangle$ satisfying
\begin{equation}
\Gamma^{\pm}_{E_1E_2} = \left[\begin{array}{cccc} N_{E_1} & 0 & C^\pm_{E_1E_2} & 0\\[.05in] 
0 & N_{E_2} & 0 & C^\pm_{E_1E_2}\\[.05in]
C^\pm_{E_1E_2} & 0 & N_{E_1} + 1 & 0 \\[.05in]
0 &  C^\pm_{E_1E_2} & 0 & N_{E_2} + 1\end{array}\right].
\end{equation}
Here, $N_{E_1} \equiv \langle \hat{a}_{E_1}^\dagger\hat{a}_{E_1}\rangle$, $N_{E_2} \equiv \langle \hat{a}^{\pm\dagger}_{E_2}\hat{a}^\pm_{E_2}\rangle$, and $C^\pm_{E_1E_2} \equiv \langle \hat{a}_{E_1}^\dagger\hat{a}^\pm_{E_2}\rangle$, where we have exploited the photon-number preserving nature of BSPK modulation and taken the cross-correlation to be real valued.  Eve's Holevo information (in bits) is 
\begin{equation}
\chi_{EB} = S(\hat{\boldsymbol\rho}_{E_1E_2}) -M [S(\hat{\rho}^+_{E_1E_2}) + S(\hat{\rho}^-_{E_1E_2})]/2,
\end{equation}
where $\hat{\boldsymbol\rho}_{E_1E_2}$ is Eve's \em unconditional\/\rm\ joint state for all $M$ mode pairs, and $S(\hat{\rho})$ is the von Neumann entropy (in bits) of the density operator $\hat{\rho}$.  The conditional density operators $\hat{\rho}^\pm_{E_1E_2}$ are zero-mean, jointly-Gaussian states whose von Neumann entropies are easily evaluated from the symplectic-decomposition technique developed in \cite{pirandola08}.  Eve's unconditional density operator is \em not\/\rm\ Gaussian, but it is zero-mean and its covariance matrix, ${\boldsymbol \Gamma}_{E_1E_2}$, is block diagonal with each block equaling 
\begin{eqnarray}
\lefteqn{(\Gamma^+_{E_1E_2} + \Gamma^-_{E_1E_2})/2 = }\\ 
&=&\left[\begin{array}{cccc} N_{E_1} & 0 & 0 & 0\\[.05in] 
0 & N_{E_2} & 0 & 0\\[.05in]
0 & 0 & N_{E_1} + 1 & 0 \\[.05in]
0 &  0 & 0 & N_{E_2} + 1\end{array}\right].
\end{eqnarray}
Thus ${\boldsymbol \Gamma}_{E_1E_2}$ is identical to the covariance matrix of $2M$ independent thermal states, half with average photon number $N_{E_1}$ and half with average photon number $N_{E_2}$.  So, because a thermal state of average photon number $N$ has the maximum von Neumann entropy of all states with that average photon number, we get the upper bound
\begin{eqnarray}
\chi_{EB}  &\le&  \chi^{\rm UB}_{EB}
\equiv M[g(N_{E_1}) + g(N_{E_2})] \nonumber \\
&& \,\, -  M[S(\hat{\rho}^+_{E_1E_2}) + S(\hat{\rho}^-_{E_1E_2})]/2
\end{eqnarray}
on Eve's Holevo information, where $g(N) \equiv (N+1)\log_2(N+1) - N\log_2(N)$ is the von Neumann entropy (in bits) for the thermal state with average photon number $N$.

\section{VII.\,\, Noise threshold for classicality}
Alice's returned and retained light are in a classical state when Bob's ASE has per-mode average photon number $N_B$ such that
\begin{equation}
|\langle \hat{a}_{I_1}\hat{a}^\pm_{S_8}\rangle|^2 \le \langle \hat{a}^\dagger_{I_1}\hat{a}_{I_1}\rangle\langle \hat{a}^{\pm\dagger}_{S_8}\hat{a}^\pm_{S_8}\rangle.
\end{equation}
Evaluating the moments in this inequality yields
\begin{equation}
\kappa_I\kappa_A'\kappa_2\kappa_B'G_B\kappa_B\kappa_1\kappa_A N_S(N_S+1) \le \kappa_IN_S(\kappa_A'\kappa_2\kappa_B'G_B\kappa_B\kappa_1\kappa_A N_S + \kappa_A'\kappa_2\kappa_B'N_B),
\end{equation}
which simplifies to
\begin{equation}
N_B \ge N_B^{\rm thresh} \equiv \kappa_1\kappa_A\kappa_BG_B.
\end{equation}

\end{document}